\documentclass[sigconf,screen]{acmart}

\usepackage{natbib}
\usepackage{booktabs}
\usepackage{enumitem}
\usepackage{tcolorbox}
\usepackage{graphicx}
\usepackage{subcaption}
\usepackage{microtype}
\usepackage{censor}
\usepackage{soul}
\usepackage{amsmath}
\usepackage{xcolor}
\usepackage{xspace}
\newtcolorbox{takeawaybox}{colback=blue!10,colbacktitle=blue!40,title=Takeaways,colframe=blue!40,coltitle=black}

\setcopyright{acmcopyright}
\copyrightyear{2024}
\acmYear{2024}
\acmDOI{XXXXXXX.XXXXXXX}

\usepackage{xcolor}

\begin{document}
\title{Exploring LLM-based Agents for Root Cause Analysis}

\author{Devjeet Roy}
\authornote{This work was done during an internship at Microsoft.}
\email{devjeet.roy@wsu.edu}
\affiliation{%
  \institution{Washington State University}
  \city{Pullman}
  \state{Washington}
  \country{USA}
  \postcode{99163}
}

\author{Xuchao Zhang}
\email{xuchaozhang@microsoft.com}
\affiliation{%
  \institution{Microsoft}
  \city{Redmond}
  \state{Washington}
  \country{USA}
  \postcode{98052}
}

\author{Rashi Bhave}
\email{     }
\affiliation{%
  \institution{Microsoft}
  \city{Bengaluru}
  \state{Karnataka}
  \country{India}
  \postcode{560001}
}

\author{Chetan Bansal}
\email{chetanb@microsoft.com}
\affiliation{%
  \institution{Microsoft}
  \city{Redmond}
  \state{Washington}
  \country{USA}
  \postcode{98052}
}

\author{Pedro Las-Casas}
\email{pedrobr@microsoft.com}
\affiliation{%
  \institution{Microsoft}
  \city{Redmond}
  \state{Washington}
  \country{USA}
  \postcode{98052}
}

\author{Rodrigo Fonseca}
\email{Fonseca.Rodrigo@microsoft.com}
\affiliation{%
  \institution{Microsoft}
  \city{Redmond}
  \state{Washington}
  \country{USA}
  \postcode{98052}
}

\author{Saravan Rajmohan}
\email{saravan.rajmohan@microsoft.com}
\affiliation{%
  \institution{Microsoft}
  \city{Redmond}
  \state{Washington}
  \country{USA}
  \postcode{98052}
}

\renewcommand{\shortauthors}{Roy et al.}
\def\tname{Azure Fundamental Team}
\def\rcaagent{\textsc{RCA Agent~}}
\newcommand{\company}{Microsoft\xspace}
\definecolor{myred}{RGB}{255,0,0}
\definecolor{mygreen}{RGB}{0,128,0}

\newcommand{\greenuparrow}{\textcolor{mygreen}{$\uparrow$}}
\newcommand{\reddownarrow}{\textcolor{myred}{$\downarrow$}}
\begin{abstract}
The growing complexity of cloud based software systems has resulted in incident management becoming an integral part of the software development lifecycle. Root cause analysis (RCA), a critical part of the incident management process, is a demanding task for on-call engineers, requiring deep domain knowledge and extensive experience with a team's specific services. Automation of RCA can result in significant savings of time, and ease the burden of incident management on on-call engineers. Recently, researchers have utilized Large Language Models (LLMs) to perform RCA, and have demonstrated promising results. 
However, these approaches are not able to dynamically collect additional diagnostic information such as incident related logs, metrics or databases, severely restricting their ability to diagnose root causes. In this work, we explore the use of LLM based agents for RCA to address this limitation. We present a thorough empirical evaluation of a \react\ agent equipped with retrieval tools, on an out-of-distribution dataset of production incidents collected at a large IT corporation. 
Results show that \react\ performs competitively with strong retrieval and reasoning baselines, but with highly increased factual accuracy. We then extend this evaluation by incorporating discussions associated with incident reports as additional inputs for the models, which surprisingly does not yield significant performance improvements. Lastly, we conduct a case study with a team at \company{} to equip the \react\ agent with tools that give it access to external diagnostic services that are used by the team for manual RCA. Our results show how agents can overcome the limitations of prior work, and practical considerations for implementing such a system in practice. 
\end{abstract}

\ccsdesc[500]{Computer systems organization~Cloud computing}
\ccsdesc[500]{Software and its engineering~Maintaining software}

\keywords{Incident Management, Cloud Computing, Root Cause Analysis, AIOps}

\def\cbleu{\textsc{C-BLEU}}
\def\sbleu{\textsc{S-BLEU}}
\def\rouge{\textsc{rougeL}}
\def\meteor{\textsc{METEOR}}
\def\bertscore{\textsc{BertS}}
\def\retrievalblong{\textsc{Retrieval Baseline (k=10)}}
\def\retrievalb{\textsc{RB (k=10)}}
\def\cot{\textsc{CoT}}
\def\ircot{\textsc{IR-CoT}}
\def\ircotbm{\textsc{IR-CoT BM25}}
\def\ircotst{\textsc{IR-CoT ST}}
\def\react{\textsc{ReAct}}
\def\reactbr{\textsc{ReAct BR}}
\def\reactsqbm{\textsc{ReAct S+Q BM25}}
\def\reactsqst{\textsc{ReAct S+Q ST}}

\maketitle
\section{Introduction}
For the last several decades, large scale enterprises have been transforming their software into cloud services. With the rise of Artificial Intelligence (AI) in recent years, there has been even greater movement of computation from consumer devices to the cloud. This shift in paradigm has brought with it complex software systems that are characterized by multi-tiered architectures, microservices and distributed applications. The increased complexity of these systems makes them highly susceptible to production incidents. When left unresolved, these incidents can incur substantial costs and disrupt critical services. Therefore, prompt mitigation and resolution of these incidents is crucial to maintaining service availability and reliability~\cite{Zeng2023-em}.  However, cloud incident management~\cite{Lou2022-yy,Chen2023-sj} is extremely labor-intensive. On-call engineers (OCEs) require extensive experience with a team's services and deep domain knowledge to be effective at incident management. Even for experienced OCEs, incident management represents a time-intensive endeavor. As software systems continue scaling in size and complexity, the demands placed on OCEs and incident management systems is only bound to increase in the future. To address these challenges, the field of AIOps (Artificial Intelligence for IT Operations) has proposed numerous techniques to ease incident management. Despite these developments, several parts of the incident management lifecycle still largely rely on human intervention.

One of the most challenging aspects of cloud incident management is root cause analysis (RCA). Before an incident can be resolved, OCEs must identify the root cause of the incident to ensure that any resolution actions comprehensively and correctly fix
the incident. RCA represents one of the most labor- and skill-intensive components of the incident management lifecycle~\cite{Ma2020-io}. Even a veteran software engineer might need to spend several years on a team before they are able to effectively perform RCA on a team's services. Therefore, it comes as no surprise that researchers have tried to automate parts of this process. Numerous techniques have been proposed to assist OCEs with RCA, such as incident prioritization and retrieval of similar historical incidents. While earlier approaches focused on automating parts of the root cause analysis process, the remarkable abilities demonstrated by Large Language Models (LLMs) in recent years has increased focus on end-to-end systems for RCA. Recently, Ahmed et al.\cite{Ahmed2023-ov} proposed the use of fine-tuned LLMs for incident root cause analysis and mitigation. They showed that LLMs can find root causes of incidents even when working with a very limited set of information about an incident. Chen et al.\cite{Chen2023-js} propose RCACopilot, which expands upon this work and add retrieval augmentation and diagnostic collection tools to the LLM-based root cause analysis pipeline. They design custom workflows for different types of incidents that trigger data collection procedures, which are then aggregated to predict a root cause category for the incident and help OCEs with root cause analysis. 

While these approaches have shown promising results on the ability of LLMs to perform RCA, neither equips the LLM to dynamically query real time diagnostic information about the service(s) affected by an incident. \textsc{RCACopilot}~\cite{Chen2023-js} relies on predefined handlers that must be engineered by hand, and predicts root cause categories rather than specific root causes, while Ahmed et al.\cite{Ahmed2023-ov} rely only on the incident title and description for predicting the root cause. What's missing here is a critical step that is taken by OCEs in real world RCA scenarios - \textit{for any incident, one of the first steps performed by OCEs is collection of novel diagnostic data that is not present in the incident report}. In prior work, LLMs do not have the ability to interact with the outside environment to be able to collect this data. In this work, we propose the use of LLM-based agents -- systems that can reason, plan and interact with the external environment to collect new information -- to address this limitation and help with root cause analysis.

Despite the remarkable capabilities demonstrated by LLM-based agents across diverse domains and tasks, adapting them for the purposes of RCA represents a significant challenge. Incident production data is highly confidential, and likely out of distribution for LLMs without fine-tuning, which can be costly and impractical for large models~\cite{Chen2023-js}. In-context examples can serve as an alternative to fine-tuning for domain adaptation, but for agent based RCA, crafting entire reasoning trajectories can be challenging. This is exacerbated by the fact that agents require sophisticated prompting and typically also require fine-tuning~\cite{Yao2022-uc} or in-context examples~\cite{Song2022-ce}. Lastly, RCA poses some unique characteristics that differentiate it from standard NLP tasks. For most NLP tasks, relevant external tools such as web search engines and document retrieval are easy to use in a single step process, and do not require much prior knowledge from the LLM. For RCA, crafting a query for search or retrieval requires much more specialized domain knowledge; many sources of information such as logs, traces, and monitoring services involve querying and processing of tabular data using specialized query languages as well as knowledge of ancillary information (e.g. which database to query). Therefore, while LLM agents offer exceptional abilities that go far beyond prior approaches, it is unclear whether they can be effectively adapted to the RCA task.

In this work, we present an empirical evaluation of an LLM-based agent, \react\ for root cause analysis for cloud incident management. Our goal is to answer two important questions in this regard: 1) Can LLM agents be effective at RCA in the \textit{absence} of fine-tuning? and 2) What are the practical considerations of using LLM agents in real world scenarios? To answer these questions, we first conduct an evaluation of the \react\ agent equipped with retrieval tools on a static dataset, mirroring the evaluation setting by Ahmed et al.~\cite{Ahmed2023-ov}. In this setting, the agent does not have access to specialized, team specific, diagnostic services, thereby restricting its abilities. This establishes a lower bound for their performance, and also reflects a practical scenario where agents are incrementally adopted across an organization or company, gradually gaining access to diagnostic services over time. Next, we investigate the use of discussion comments from historical incident reports to augment our retrieval corpus. This serves two purposes; not only do discussion comments add additional context to the incident report, but they also contain records of the diagnostic steps followed by OCEs for past incidents. The latter can potentially be used in lieu of few-shot examples to guide the agent. Lastly, to explore the full potential of agents, we present a case study of a practical implementation of an LLM agent for RCA, fully equipped with team specific diagnostic resources, in collaboration with another team at \company . Concretely, we make the following contributions:
\begin{itemize}[leftmargin=*]
    \item We present the first empirical study on the use of \react~\cite{Yao2022-uc}, an LLM agent, for RCA in an out of domain setting on a static dataset of real world production incidents
    \item We conduct a qualitative analysis of the different success and failure modes of the \react\ in RCA. 
    \item We evaluate the use of discussion comments from historical incidents and its impact on the agent's performance. 
    \item We present a case study of a real world implementation of an LLM-based agent for RCA with a team at a large scale enterprise
    \item We highlight both the potential of LLM-based agents and the challenges involved in implementing real world systems capable of fully autonomous RCA.
\end{itemize}  

\section{Background and Related Work}

\subsection{Cloud Incident Management and Root Cause Analysis}

Production incidents are unplanned events or disruptions in service that adversely affect customers. Outages in service due to production incidents can be extremely costly for enterprises. The complexity of modern software systems renders production incidents inevitable, and \textit{incident management} a key component of the software development life cycle. The life cycle of an incident involves incident detection, triaging, diagnosis and mitigation~\cite{Ahmed2023-ov}. While incidents may be reported by customers or automatically detected and triaged using monitoring services, the remaining steps are traditionally conducted by one or more on-call engineers (OCEs). The goal of incident management is to minimize the time between the occurrence of the incident, and its resolution. 

\subsection{Root Cause Analysis (RCA)}

\begin{figure}[h]
\begin{tcolorbox}[arc=0pt,outer arc=0pt, ,nobeforeafter,colback=blue!0,boxrule=1pt,colframe=black!80, left=1mm,
  right=1mm,
  top=1mm,
  bottom=1mm,]
    
\textbf{Title: } SD\#1234123412341234 | PRE | SEV A | Specified blob does not exist. | Cloud Services LLC 
\\
\textbf{Description: } 
Customer mentioned that after stopping stream analytics on 09/23 they are getting errors on streaming into <\emph{database product}>%
[\textbf{...}]
It was throwing an error "Specified blob does not exist” and “Invalid connection string format. [SessionID: <\emph{uuid}> %
Found Another error message "Error while Ingesting data to <\emph{database product}>"%
\end{tcolorbox}
\caption{Example Incident}\label{fig:example_incident}

\end{figure}

\textit{Root Cause Analysis} constitutes one of the most time-consuming aspects of the incident management life cycle. When OCEs receive an incident, they systematically perform a series of troubleshooting steps to identify the root cause. Each troubleshooting step yields previously unknown information, helping the OCE narrow down on the set of plausible root causes. 
This highlights a key aspect of root cause analysis: the process of collecting additional diagnostic information related to the incident. The incident report describes the symptoms leading to the reporting of the incident, but similar symptoms can emerge from distinct root causes, which might span a diverse set of domains, such as hardware failures, network issues or software bugs.
Therefore, OCEs must start the diagnosis process by collecting supplementary data from relevant logs, metrics and other monitoring and diagnostic services. 
For example, the incident shown in Figure~\ref{fig:example_incident} was resolved by checking logs collected from the affected service to identify the sequence of events that lead to the failure encountered by the customer. 
Another implicit requirement in this process is that OCEs know 1) what additional information needs to be collected, and 2) how to collect this information.
This is why even experienced engineers need to have experience with team's services before they can effectively perform RCA. In all, successful RCA requires the following pieces of information: 1) symptoms reported in the incident report, 2) additional diagnostic information, and 3) domain expertise, i.e. what diagnostic information should be collected based on the information, how to collect it and general knowledge about the application domain

The root cause analysis pipeline demonstrates many of the challenges posed for OCEs as well as efforts to automate this procedure. OCEs must have sufficient domain knowledge and familiarity with the affected service to know 1) which supplementary data to collect, 2) how this data must be collected and 3) how to analyze all of the available information (including the incident report). Depending on the scale and complexity of the underlying service, this might require OCEs to have several years of experience with the team's services to develop the requisite skill set for effective root cause analysis. Even when OCEs are sufficiently trained, the data collected can be multi-faceted, spanning from structured tabular data to unstructured logs and customer reports. This further complicates data analysis and subsequent hypothesis generation for OCEs. While OCEs can overcome these challenges by leveraging domain expertise and experience, this poses a significant challenge for prior automated approaches, that are unable to collect this supplementary data, let alone analyze it to produce a root cause.  

\subsection{Automated RCA}

Numerous studies have proposed various techniques for automating root cause analysis, such as using machine learning models and deep learning models ~\cite{Soldani22-hg} to identify patterns in event data and determine the underlying causes of incidents. Another important area of research in RCA is the use of anomaly detection models ~\cite{Soualhia22-ev}, such as statistical, machine learning and deep learning models ~\cite{Hagemann21-an}, have been proposed to identify anomalies in system behavior and alert operators in real-time. 
Studies have proposed various techniques for RCA and triage such as learning a hierarchical monitoring system ~\cite{Nair15}, diagnosing and triaging performance issues ~\cite{Chetan19-dc}, and correlating events with time series ~\cite{Luo14-en}. In addition, there have been studies exploring the use of structured knowledge mining from various artifacts, such as incident reports and root cause documentation, to mine structured knowledge in software engineering such as troubleshooting guides (TSGs) ~\cite{Jiang22-eg} and there have been efforts to improve TSG quality ~\cite{Shety22-lr}and make them more effective for incident resolution.

Large Language Models (LLMs) have shown remarkable ability to work with a wide variety of data modalities, including unstructured natural language, tabular data and even images. 
Recently, Ahmed et al.~\cite{Ahmed2023-ov} proposed the use of fine-tuned pretrained LLMs for RCA of cloud incidents. Since incident data is highly confidential, and unlikely to have been observed by pretrained LLMs, fine-tuning is necessary for domain adaptation of vanilla LLMs. In this work, we adopt the RCA task as framed in Ahmed et al.~\cite{Ahmed2023-ov}; given an incident report, we want our model to predict a specific root cause. However, unlike the original setting, we exclude the use of fine-tuning or other training approaches for domain adaption. As pointed out by Chen et al.~\cite{Chen2023-js}, while fine-tuning can be effective, it is also costly and time-consuming, and must be repeated every time the base model gets updated, or services evolve. 
To address these limitations, Chen et al.~\cite{Chen2023-js} introduce \textsc{RCACopilot}, which uses predefined handlers to automatically collect multi-modal diagnostic data relevant to the incident, and an LLM to analyze the collected data and predict a root cause category for the incident that serves to assist OCEs with RCA, without the need for finetuning.
Unlike \textsc{RCACopilot}, the \react\ agent presented in our case study can dynamically collect related diagnostic data autonomously, without the need for predefined handlers.

\subsection{Augmented LLMs and LLM-Based Agents}
A recent development in LM research has been the rise of LMs augmented with the ability to reason and use tools, or \textit{Augmented Language Models (ALMs)}~\cite{Lewis2020-rj,Mialon2023-rn,Schick2023-vu}. 
Augmenting LLMs extends their ability beyond what is possible in a purely language modelling regime. Primarily, these augmentations are either external components that allow the LLM to interact dynamically with its environment for a given problem setting, or prompting techniques that endow the LLM with sophisticated reasoning abilities for complex analytical tasks~\cite{Wei2022-vl}. For example, LLMs have been augmented with external retrieval databases that can factually ground their predictions, as well as allow them to use information that was not seen in training. Retrieval can also narrow the gap between smaller models and their larger counterparts. LLMs can also be augmented with external components beyond retrieval, such as code interpreters~\cite{noauthor_undated-eg} and web search engines. 
More recently, LLM-based agents combine the external augmentation components with reasoning and planning abilities to allow the LLM to autonomously solve for complex tasks such as sequential decision-making problems~\cite{Shinn_undated-kf}, knowledge-intensive question answering~\cite{Trivedi2022-pf} and self debugging~\cite{Chen2023-hj}.

\section{LLM-Based Agents for RCA}

An LLM agent is an ALM that has the ability to both reason and use tools. In recent years, several different formulations of LLM agents have been proposed~\cite{Yao2022-uc,Song2022-ce}. For this work, we base the \rcaagent on the \textsc{ReAct} framework~\cite{Yao2022-uc}.  This framework interleaves reasoning and tool usage steps, combining principles from reasoning-based approaches such as \textsc{Chain of Thought}~\cite{Wei2022-mi} with tool usage models like \textsc{Toolformer}\cite{Lewis2019-hp}. \textsc{ReAct} is a natural fit for the RCA task for many reasons: 1) Real-world RCA task has elements of both sequential decision-making (deciding which troubleshooting steps to take) and knowledge-intensive question answering (assessing available diagnostic information to produce a candidate root cause), both of which are supported by \react; 2) in an out of distribution setting such as the one we consider, \react can quickly adapt to new information since it interleaves reasoning, planning and environment feedback rather than creating a long-horizon plan upfront; and 3) it can easily be augmented with additional components such as reflection~\cite{Shinn2023-hb} and external memory mechanisms~\cite{Zhao2023-rd} which would benefit RCA for incidents requiring a longer diagnostic process.

\subsection{Overview}
\begin{figure}[h]
\includegraphics[width=0.5\textwidth]{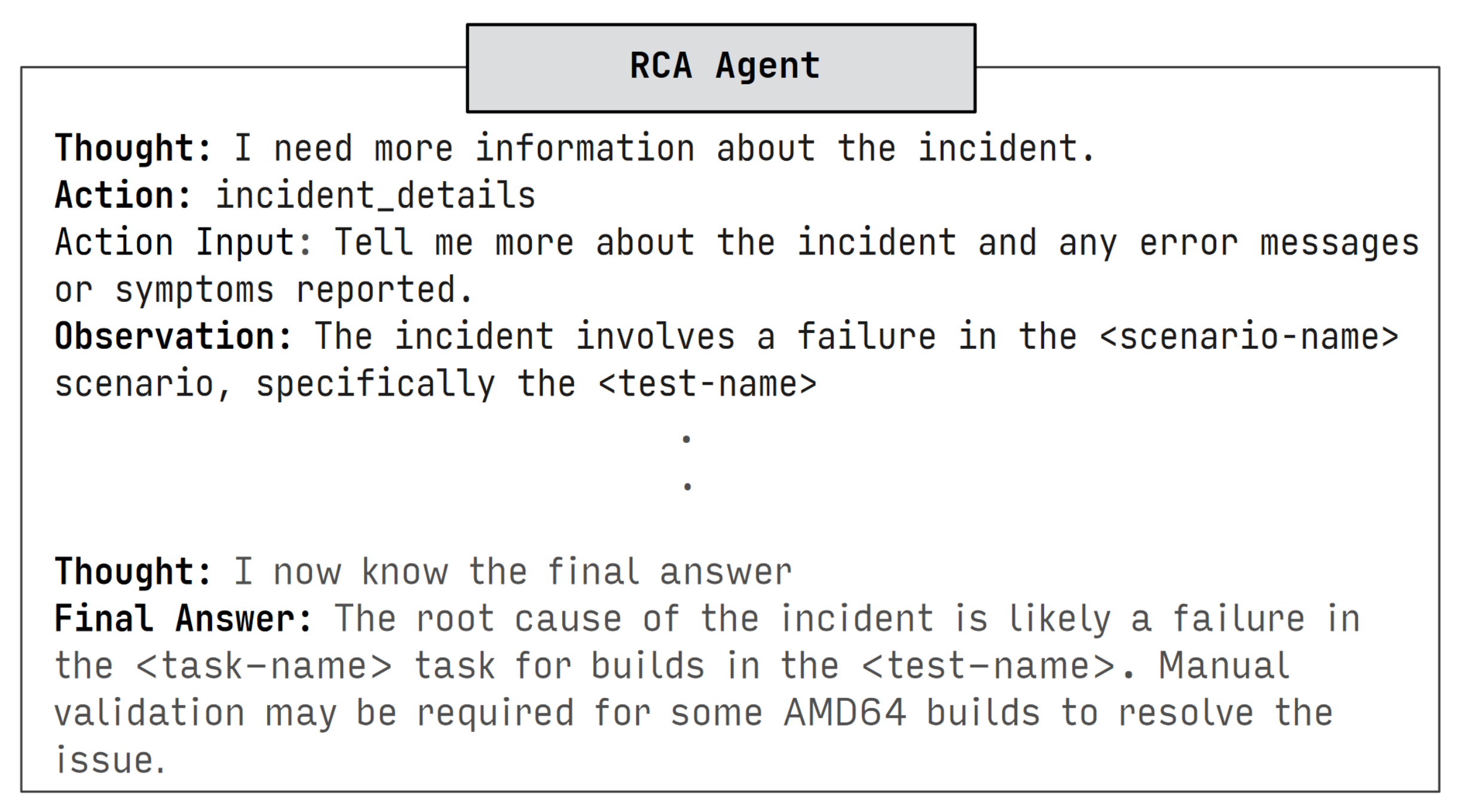}
\caption{An example of \react's reasoning trajectory}
\label{fig:sample_react_trajectory}
\end{figure}
Figure~\ref{fig:sample_react_trajectory} shows an example of a sample trajectory produced by a \react\ agent: the agent produces a "thought", or a reasoning step that informs the next "action" it takes. The action space is consists of a fixed set of tools available to the agent. Once the action and it's inputs are specified, the tool is executed and it's outputs are reported back to the agent as an observation. 
Steps 1-3 repeat for as many times as needed to perform the task at hand. We use the Langchain\cite{Chase2022-sg} framework to implement the \react\ agent. Note that the tools used by the agent might also rely on LLMs. To disambiguate, we refer to the LLM performing the \react\ loop as the \textit{planner}. We limit the maximum number of iterations of this loop to 20 due to time and resource constraints. 

\subsection{Zero-Shot Prompting}

While LLM-based agent approaches typically benefit with few-shot examples~\cite{Yao2022-uc, Song2022-ce}, we use \react\ in a much more challenging setup with a zero-shot prompt. Originally, we set out to craft few-shot examples based on examples from the evaluation set. However, for the setting in RQ1 and RQ2, where we only utilize the incident title and descrption, we found it extremely challenging to come up with reasoning traces grounded in the available information that would arrive at the correct root cause.

\subsection{Agent Evaluation}
 One of the primary benefits of agent-based RCA is their ability to collect external diagnostic information via tools. This is difficult to evaluate without the existence of a simulated environment such as \textsc{WebArena} ~\cite{Zhou2023-ds}, \textsc{AlfWorld}~\cite{Shridhar2020-ky} or \textsc{WebShop}~\cite{Yao2022-hq}. The main challenge in constructing such an environment is that it is difficult to determine what diagnostic services were used to diagnose a particular incident, since OCEs are not required to report each and every diagnostic step taken. Moreover, the type of diagnostic services used by different teams can vary greatly. Another challenge is that the environment needs to support not only the most optimal troubleshooting trajectory that the agent can take, but a reasonably large subset of other plausible trajectories, i.e. even if we know what diagnostic data is needed to resolve an incident, it does not suffice to only capture this specific data for the environment. Given such an evaluation environment does not currently exist, we evaluate the agent in a restricted setting where we do not assume access to any specialized services, similar to \cite{Ahmed2023-ov}. 
 While this evaluation does not reflect the benefits of the agent's ability to perform autonomous diagnostic steps, it provides us with a lower bound for performance of the agent when specialized tools are unavailable, and allows us to fairly compare it to other ALMs that do not have the ability to query additional diagnostic data. In addition, to demonstrate the agent's ability to interact with diagnostic services, we also present a case study of a prototype implementation of an agent in collaboration with a team at our company. This presents a more realistic evaluation of the agent but at a much smaller scale. The goal of the case study is to examine the benefits and limitations of the agent in a practical environment, and to identify practical considerations for real world adoption based.

\subsection{Tools}
For the generalized setting, we restrict ourselves to general tools that apply to \textit{all} incidents, regardless of their place of origin.

\noindent\textbf{Incident Details} Our investigation of the evaluation dataset revealed that sometimes incident reports contain logs, stacktraces and other diagnostic information. While this information can be noisy and present challenges with context length, it is possible to expose the raw incident description to the LLM via a question-answering tool. The agent can use this tool to answer specific questions about the incident that might get lost during summarization. 

\noindent\textbf{Historical Incidents} This tool retrieves historical incidents based on the query made by the LLM planner. Based on experiments on our development dataset, we formulate two variants of this tool. The first variant uses the target incident title and description, along with a query produced by the agent for retrieval, and simply returns the retrieved documents as an observation without further processing. Since the query here is a passage, we exclusively use the SentenceTransformer retriever for retrieval.
We refer to \react\ agents using this variant of the tool as \reactbr.
The second variant uses utilizes a two-step retrieval process. The LLM Planner must first generate a query to search for historical incidents. Then, the planner can perform question-answering over the retrieved set of historical incidents. The tool uses an LLM injected with the retrieved incidents to answer the planner's question. 
This two step process allows the planner to disentangle the retrieval query from the target incident report, and also mitigates instances where the size of retrieved historical incidents might extend beyond the context length of the underlying LLM. We restrict the retrieval tool to retrieve $k=3$ documents per query, to give the agent the opportunity to create a diverse set of queries while still maintaining an overall budget of 10 retrieved documents for parity with other baselines.

\section{Research Questions}

To evaluate the efficacy of LLM-based Agents in RCA, we ask the following research questions:

\noindent\textbf{RQ1: How effective are LLM-based agents at finding incident root causes when given access to a generalized toolkit?}
\noindent In this setting, we test the efficacy of LLM-based agents at root cause analysis in an out of distribution setting when they are given access only to tools that are independent of specific teams. We equip the agent with a generalized retrieval tool over historical incidents, and a question-answering tool over the raw incident description. We consider various strong ALM baselines that, unlike the agent, are unable to use tools.

\noindent\textbf{RQ2: Do discussion comments help improve LLM based approaches to root cause analysis?}

\noindent Discussion comments on incident reports contain records of the diagnostic steps taken by OCEs to resolve the incident, and can guide models in performing RCA on future incidents. Here, we aim to investigate whether incorporating these discussion comments into our retrieval corpus of historical incidents impacts the performance of the agent as well as selected baselines from RQ1. To perform this evaluation, we augment incidents in our retrieval corpus with associated discussion comments \textit{post-retrieval}, to ensure that the presence of the comments does not affect retrieval. 

\noindent\textbf{RQ3: How effective are LLM based agents at root cause analysis when given access to specialized tools used by a team for incident management?}

\noindent In this research question, we evaluate a real world scenario when an LLM based agent has access to a team specific knowledge base and monitoring service. To conduct this evaluation, we perform a case study with another team's on-call engineers. We package the \react\ agent with these resources into a chat interface, and conduct an in person experiment to see if this agent is able to effectively assist the on call engineer in finding the root cause of a small set of incidents. 

\section{Methodology}
We describe the methodology used to answer \textsc{RQ1} and \textsc{RQ2} in this section. The methodology for \textsc{RQ3} is described in Section~\ref{sec:case_study}.
\subsection{Dataset}
We collect incident data from our internal incident portal, from 01/01/2020 to 09/30/2021. Our data collection process yielded a total of 107,000 unique incidents, which we split into a train (102,000), evaluation (2000) and test (3000) sets. For this work, we randomly sample 100 incidents from the evaluation set and 500 incidents from the test set to reduce costs, in line with work in NLP~\cite{Trivedi2022-tm}. We use  the training set is primarily used as the retrieval corpus for our experiments. Like Ahmed et al.~\cite{Ahmed2023-ov}, we use the incident title and description as the primary sources of information about the incident. For RQ2, we also include discussion comments into the historical corpus. Incident descriptions and root causes do not follow a standard format, and can be quite long. This imposes limitations on the number of historical incidents that can be fit in context when using any kind of retrieval augmented generation. Hence, we use \textsc{gpt-3.5-turbo} to summarize descriptions and root causes. For RQ2, we also summarize discussions comments. Since discussion comments are much longer, we split them into chunks, summarize each individual chunk and recombine them, utilizing the LLM for each step. Note that the summarization process is difficult to evaluate due to a lack of reference summaries, and hence we rely on qualitative analysis and end-to-end evaluation on RCA to iterate on the summarization process.   

\subsection{Base LLMs}
For all of our experiments, we use OpenAI GPT4-8k~\cite{OpenAI2023-la} as the primary language model. GPT-4 is the most powerful model in OpenAI's repository of models, and is one of the few models that can be used to reliably drive an agent in a zero-shot setting. The large context size (8,000 tokens) also enables us to use a larger number of retrieved incidents for our models. For summarization of incidents and discussion comments, we use \textsc{gpt-3.5-turbo} to lower costs.

\subsection{Retrievers}
We construct a retrieval corpus of historical incidents that encompasses the entire training split of our collected dataset. We consider one dense retriever and one sparse retriever.

\noindent\textbf{Dense Retriever (ST)} We use a pretrained Sentence-Bert~\cite{Reimers2019-pg} based encoder (\textsc{all-mpnet-base-v2}) from the associated \textsc{SentenceTransformers} as our dense retriever and Max Marginal Relevance (MMR)~\cite{Carbonell1998-pu} for search.

\noindent\textbf{Sparse Retriever (BM-25)} While models that perform a single retrieval step, other models such \textsc{IR-CoT} and the \textsc{ReAct} agent perform multiple retrieval steps with different queries, and can benefit from term based search~\cite{Trivedi2022-tm}. We use \textsc{BM-25}~\cite{Robertson2009-nl} as our sparse retriever. 

\subsection{Baseline Models}
Here, we describe the baselines used for our evaluation in RQ1 and RQ2. We restrict ourselves to ALMs that do not require any fine-tuning. All the following baselines use historical incident retrieval, and are restricted to a retrieval budget of $k=10$.

\noindent\textbf{\textsc{Retrieval Baseline (RB)}} 
Retrieval Augmented Generation (RAG) is an effective strategy to providing domain adaptation for language models without additional training. For our experiments, we create a retrieval database of historical incident reports with known root causes, and use the incoming incident's title and description to retrieve top-k relevant historical incidents. These incidents are then put into the LLM's context as few-shot examples. 

\noindent\textbf{Chain of Thought (\textsc{CoT})} Chain of Thought is one of the earlier prompting methodologies developed to enhance the reasoning abilities of LLMs\cite{Wei2022-vl}. The idea here is to encourage the model to break the input problem into smaller parts by thinking step by step. For our experiments, we use CoT in a zero-shot setting, by appending a prefix ("Let's think step by step") to the answer prompt.

\noindent\textbf{Interleaving Retrieval - Chain of Thought (\textsc{IR-CoT})} Trivedi et al.\cite{Trivedi2022-pf} show that interleaving vanilla \textsc{CoT} prompting with retrieval improves model performance on complex, multistep reasoning tasks. After every reasoning step the LLM takes, the reasoning step is used to retrieve relevant documents from the retrieval corpus. This is shown to improve performance over using single step retrieval for knowledge intensive question answering tasks.

\subsection{Automatic Evaluation Metrics}

For evaluating models in the general setting, we use a 3 evaluation metrics based on lexical similarity (BLEU, METEOR, Rouge) and 1 on semantic similarity (\bertscore). \textsc{BLEU}~\cite{Papineni2002-yq} is a precision based lexical similarity metric that computes the n-gram overlap between model predictions and ground truth references. We use both corpus (\textsc{C-BLEU}) and segment (\textsc{S-BLEU}) level variants. \meteor~\cite{Banerjee2005-cp} considers both precision and recall, and uses more sophisticated text processing and scoring systems. \rouge~\cite{Lin2004-sf} is commonly used to evaluate summarization and is recall based. \textsc{BERTScore}~(\bertscore)~\cite{Zhang2019-ju} measures semantic similarity rather using pretrained BERT models. 

\subsection{Qualitative Analysis}

\begin{table}[!htb]
\centering
\caption{Manual Annotation Criteria}
\label{tab:qa_rubric}
\resizebox{0.95\columnwidth}{!}{
\begin{tabular}{@{}ll@{}}
\toprule
\textbf{Outcome}           & \textbf{Description}                                                        \\ \midrule
\textbf{Correct} \\
Precise                 & Precisely matches reference root cause                                      \\
Imprecise               & Matches reference but misses some details                        \\
Hallucination & Matches reference but contains unrelated factual errors          \\\midrule
\textbf{Incorrect} \\
Hallucination           & Contains factual errors in reasoning or prediction                          \\
Insufficient Evidence   & Refrains from making a prediction                                           \\
Other                   & Cause of error unknown                                                      \\
Reasoning Error         & Reasoning contains errors                                                   \\
Retrieval Error         & Unable to retrieve relevant historical incidents \\ \bottomrule
\end{tabular}
}
\end{table}
While automatic metrics can serve as proxies for lexical and semantic similarity, they are not able to accurately measure factual accuracy or conclusively establish semantic equivalence. Common failure modes of these metrics include predictions that restate the incident report or highly generic predictions (e.g. "there was a transient network issue")~\cite{Ahmed2023-ov}, both of which can trivially boost lexical similarity. To better characterize the performance of the LLM agent and other baselines, two authors conduct a qualitative coding on a sample of 100 predictions for three models (300 annotations in total). The labelling is done in iteratively, and the authors engaged in extended discussions to resolve disagreements. We characterize both success and failure modes of these models based on the coding scheme shown in Table~\ref{tab:qa_rubric}. The coding scheme is adapted from Yao~\textit{et al.}~\cite{Yao2022-uc} and specialized for the RCA task. The adaptations are a superset of the original categories and were made after performing labelling on a smaller sample 20 predictions to distinguish useful scenarios for RCA. Notably, for correct predictions, we differentiate correct predictions that unambiguously match the reference (\textit{Precise}), match the reference semantically but exclude some specifics present in the reference (\textit{Imprecise}), and those that match the root cause semantically but also contain unrelated factual accuracies (\textit{Hallucinations}). The last case commonly manifests as predictions that suggest the execution of post-hoc resolutions actions (e.g. the incident was resolved by restarting the affected cluster) that did not take place.  \textit{Imprecise} predictions can be useful for OCEs, whereas factual errors  can mislead OCEs. For predictions that don't match the reference root cause, we add two new categories to the ones from ~\cite{Yao2022-uc}. The first, \textit{Insufficient Evidence}, refers to an incorrect prediction that indicates that there isn't enough evidence available to determine the root cause for the incident. The second, \textit{Other}, refers to instances of incorrect predictions that do not have a clearly identifiable cause for error. This is an extension to the label ambiguity category from ~\cite{Yao2022-uc}, and now includes other failure cases where the model predicts a plausible specific root cause (unlike \textit{Insufficient Evidence} which is only applied to cases where no specific root cause is indicated), but does not contain obvious reasoning, retrieval, or factual errors. This is often due to the information sparsity of incident reports, especially in cases where the incident report provides details as external links that are inaccessible for the models.

\section{\textsc{RQ1} and \textsc{RQ2} Results}

\begin{table}[!htb]
\centering
\caption{RCA performance on test set}
\resizebox{0.99\columnwidth}{!}{
\begin{tabular}{@{}lrrrrr@{}}
\toprule
\textbf{Model}             & \multicolumn{1}{l}{\textbf{\cbleu}} & \multicolumn{1}{l}{\textbf{\sbleu}} & \multicolumn{1}{l}{\textbf{\rouge}} & \multicolumn{1}{l}{\textbf{\meteor}} & \multicolumn{1}{l}{\textbf{\bertscore}} \\ \midrule
\textsc{RB} (k=3)  & 4.73                                & 4.64                                & 18.48                             & 21.62                             & 0.863                                  \\
\textsc{RB} (k=6)  & 5.66                                & 5.56                                & 19.78                             & 23.25                             & 0.865                                  \\
\retrievalb\ & 5.97                                & \textbf{5.74}                                & \textbf{20.30}                    & \textbf{24.11}                    & \textbf{0.866}                         \\ \midrule
\cot                        & \textbf{6.31}                       & 5.60                                & 19.91                             & 22.02                             & 0.865                                  \\ \midrule
\ircotst                 & 3.91                                & 3.67                                & 16.97                             & 18.50                             & 0.859                                  \\
\ircotbm               & 4.61                                & 4.02                                & 17.56                             & 19.94                             & 0.860                                  \\ \midrule
\reactbr                 & 5.53                                & 4.90                                & 17.45                             & 19.23                             & 0.858                                  \\
\reactsqbm             & 5.59                                & 4.73                                & 17.43                             & 18.72                             & 0.857                                  \\
\reactsqst               & 5.27                                & 4.58                                & 17.35                             & 18.60                             & 0.857                                  \\ \bottomrule
\end{tabular}
}
\label{rq1_testset_results}
\end{table}

\subsection{RQ1: How effective are LLM based agents at finding incident root causes when given access to a generalized toolkit?}

Table~\ref{rq1_testset_results} presents the results for our quantitative evaluation based on automatic evaluation metrics. For the Retrieval Baseline model, we see that the number of historical retrieved has a positive impact on performance across the 4 lexical metrics. However, the impact on semantic metrics remains small ($< 1$). This is likely due to the increase in access to domain specific terms with a larger number of historical incidents. We observed similar trends in experiments on our development set, and consequently we limit $k=10$ for the remaining models. CoT offers the highest performance on \cbleu\ (6.31), followed closely by the \retrievalb (5.97), \textsc{Retrieval Baseline (k=6)}~ (5.66), \reactbr\ (5.53) and \reactsqbm\ (5.59). However, these differences in performance are well within the margin of error for \cbleu\cite{Mathur2020-ec}. Both variants of \ircot\ lag behind, performing more at a similar level as the \textsc{Retrieval Baseline (k=3)} despite retrieving a larger number of historical incidents. For the remaining lexical metrics, \retrievalb\ offers the highest levels of performance, closely followed by \cot and \textsc{Retrieval Baseline (k=6)}, with the largest difference between these models on \meteor (2.09). These are followed \textsc{ReAct} variants, \reactbr\ and \reactbr, and \textsc{Retrieval Baseline (k=3)}. When investigating reasoning logs, we discover that the \textsc{ReAct} variants retrieve a mean of 4 unique historical incidents on average, with an average of 2 lookups per incident. While on each retrieval step they retrieve 3 unique incidents, the \textsc{historical\_incidents} tool is stateless, and does not take into consideration documents that have already been retrieved in prior steps, resulting in some duplication. This is a consequence of the limited information available in the incident title and description, making it difficult for the model to crafting separate queries that yield distinct sets of historical incidents. This is likely also why the \ircot\ variants perform poorly on lexical metrics. When we consider semantic similarity, we observe a performance envelope of $< 1$ across all models. Therefore, neither reasoning nor additional historical incidents drastically change the semantic content of predictions made by these models.

\textbf{Qualitative Analysis} Table~\ref{tab:qa_results} shows the results of our qualitative assessment for the \retrievalb, \cot\ and a variant for \react(\reactsqbm). \cot\ and \retrievalb\ have an accuracy of 39\%, followed by \reactsqbm at 35\%. There are 28/97 examples that are solved correctly by all three models. \reactsqbm\ correctly predicts 4 examples that are incorrectly predicted by the other two. When we examine these instances, we discover that in all of these instances, \react\ was able correctly filter out (in its reasoning steps) historical incidents that share some lexical similarity with the target incident report but ultimately are semantically quite different, whereas the other two models incorrectly include them in consideration for their final prediction. \cot\ and \retrievalb\ correctly predict 8 and 9 examples respectively for which \react\ is incorrect. 2 of these instances resulted from reasoning errors by \react\, and the rest were primarily instances where it indicated a lack of evidence (\textit{Insufficient Evidence}) for RCA. Looking more closely, 26\% (10/38) of the correct predictions made by the \retrievalb\ contain hallucinations, while it is $< 1\%$ for \cot\ and \reactsqbm.  Similarly, 49\% (29/59) of the \retrievalb's incorrect predictions are hallucinations, dropping to 18\% (11/59) for \cot\ and 6\%(4/63) for \react. Overall, \react\ has the highest precision among the 3, but this comes at the cost of lower overall accuracy.

\begin{table}[!htb]
\centering
\caption{Manual Labelling of Success and Failure Cases}
\label{tab:qa_results}
\resizebox{0.9\columnwidth}{!}{
\begin{tabular}{@{}llrrr@{}}
\toprule
                   & \textbf{Type}           & \multicolumn{1}{l}{\textbf{\retrievalb}} & \multicolumn{1}{l}{\textbf{CoT}} & \multicolumn{1}{l}{\textbf{ReAct-BM25}} \\ \midrule
\textbf{Correct}   & Imprecise               & 2                                               & 7                                & 5                                       \\
                   & Hallucination & 10                                              & 1                                & -                                       \\
                   & Precise                 & 26                                              & 30                               & 29                                      \\
                   & \textbf{All}            & \textbf{38}                                     & \textbf{38}                      & \textbf{34}                             \\ \midrule
\textbf{Incorrect} & Hallucination           & 29                                              & 11                               & 4                                       \\
                   & Insufficient Evidence     & 11                                              & 19                               & 39                                      \\
                   & Other                   & 19                                              & 27                               & 8                                       \\
                   & Reasoning Error         & -                                               & 2                                & 10                                      \\
                   & Retrieval Error         & -                                               & -                                & 2                                       \\
                   & \textbf{All}            & \textbf{59}                                     & \textbf{59}                      & \textbf{63}                             \\ \bottomrule
\end{tabular}}
\end{table}

\cot\ makes fewer reasoning errors than \react. This is likely in part due to the more sophisticated prompting involved with \react\ and the zero-shot setting. We observed some instances of longer reasoning trajectories wherein \react\ would have difficulty maintaining the prompt format. 66\% of \react's incorrect predictions indicate lack of information to make a root cause prediction (\textit{Insufficient Information}, while this is much less frequent for \cot\ (32\%) and \retrievalb\ (18\%). Lastly, a notable portion of errors for the \retrievalb\ (32\%) and \cot\ (45\%) do not have a clear cause for the error (\textit{Other}), while this happens much less for \react\ (12\%). Many of these uncategorized errors are predictions that are too generic (e.g. suggesting a non-specific configuration issue), while others are plausible based on historical incidents but incorrect. 

Overall, the qualitative analysis indicates that the higher correctness rates of the \retrievalb\ come at the cost of factual accuracy, despite the grounding offered by retrieval. \cot\ offers the same correctness rate with lower rates of hallucination. 
This clearly demonstrates the benefits of introducing explicit reasoning into an LLM; both approaches utilize identical retrieval mechanisms, and the main difference is the zero-shot reasoning in \cot.
The \react\ agent also benefits from reasoning, offering the lowest rates of hallucinations for both correct and incorrect predictions, albeit at a slightly lower overall accuracy rate.
\begin{tcolorbox}[left=0mm,
  right=0mm,
  top=0mm,
  bottom=0mm,]
\textbf{RQ1 Takeaways: }\react\ agents perform competitively with retrieval and chain of thought baselines on semantic similarity, while under performing on lexical metrics. Manual labelling reveals that they achieve competitive correctness rates (35\% for \reactsqbm\ vs 39\% for the baselines), while providing a substantially lower rate of hallucinations (4\% for \react\ vs 12\% for \cot\ and 40\% for \retrievalb). 
\end{tcolorbox}

\subsection{RQ2: Do discussion comments help improve LLM based approaches to root cause analysis?}

\begin{table}[!htb]
\caption{Test set results after incorporating discussions 
}

\resizebox{0.97\columnwidth}{!}{%
    \begin{tabular}{@{}lrrrrr@{}}
    \toprule
    \textbf{Model}             & \multicolumn{1}{l}{\textbf{\cbleu}} & \multicolumn{1}{l}{\textbf{\sbleu}} & \multicolumn{1}{l}{\textbf{\rouge}} & \multicolumn{1}{l}{\textbf{\meteor}} & \multicolumn{1}{l}{\textbf{\bertscore}} \\ \midrule
    \retrievalb & \textbf{6.65 \greenuparrow}                       & \textbf{6.01 \greenuparrow}                       & \textbf{20.8 \greenuparrow}                       & \textbf{23.81 \reddownarrow}                    & \textbf{0.867}                  \\
    \cot                        & 6.18 \reddownarrow                               & 5.21 \reddownarrow                                & 18.8 \reddownarrow                              & 21.32 \reddownarrow                             & 0.861                           \\ \midrule
    \reactbr                   & 5.44 \reddownarrow                                   & 4.91                                   & 17.8 \greenuparrow                                   & 20.04 \greenuparrow                                   & 0.854                                      \\
    \reactsqbm             & 5.52                                & 4.68                                & 17.4                               & 18.96 \greenuparrow                               & 0.858                           \\ \bottomrule
    \end{tabular}
}
\label{rq2_testset_results}
\end{table}

Table~\ref{rq2_testset_results} shows the performance of the considered models after incorporating discussions into retrieved historical incidents. 
In general, incorporating discussions provides mixed results on model performance for lexical metrics across different models. 
Discussions improve performance on \cbleu, \sbleu\ and \rouge\ for \retrievalb\, but these improvements are modest. 
On the other hand, it experiences a modest drop in performance for \meteor\ ($<1$). \cot\ experiences performance degradation for all lexical metrics: \cbleu\ (-0.13), \sbleu\ (-0.39), and \meteor\ (-0.7) and \rouge\ (-1). Unlike \cot\ both \react\ variants are mostly positively impacted by the inclusion of discussions. \reactbr\ shows improvements in performance for \rouge\ (+0.1) and \meteor\ (+0.8) but obtains lower \cbleu\ (-0.1), and does not show any difference in \sbleu. Similarly, \reactsqbm\  does not display differences in \cbleu, \sbleu\ or \rouge, but gets a small improvement in \meteor\ (+0.24). It is important to note that these small differences in metric scores ($ \leq 1$) are likely not be perceivable to human annotators, as has been empirically observed in NLP~\cite{Mathur2020-ec}, as well as SE~\cite{Roy2021-ya}. Lastly, semantic metrics reveal that the incorporation of discussions does not significantly impact the performance of the 4 models in Table~\ref{rq2_testset_results}. Our qualitative observations of a small set of model predictions (20) in the presence vs absence of discussions are in line with these findings. Notably, we do not observe any meaningful differences in the semantic content of the produced root cause between the two scenarios among the predictions that we analyzed.   

We conjecture that the small observed effect of discussions on RCA performance is due to a combination of 3 factors. Firstly, comments reporting the end result of a diagnostic step constitute a large portion of RCA relevant discussions. While these comments shed light on the troubleshooting steps that lead to incident RCA and resolution, these steps cannot be \textit{replicated} by models in the general setting; models do not have access to the same diagnostic services and resources that were utilized by OCEs to arrive at the conclusions indicated by these discussion comments. Secondly, the sparsity of information present in incident titles and descriptions negatively impact the ability of models to connect information arising from discussions to the target incident. For example, a discussion comment might signal that the presence of a certain symptom indicates a particular root cause, but this symptom might not be reported in the target incident. This is especially true for symptoms that must be elicited using troubleshooting steps. Lastly, discussion threads on incident reports can themselves be quite data sparse, containing lots of administrative content that are not directly useful for RCA (e.g. incident acknowledgement, status updates). While we use length heuristics to remove clearly uninformative comments, it is inevitable that many of these low quality comments will make it past the filter unless we use more sophisticated filtering techniques (such as LLM based filtering). 
\begin{tcolorbox}[left=2mm,
  right=2mm,
  top=1mm,
  bottom=1mm,]
\textbf{RQ2 Takeaways: }Incorporating discussion comments into the historical corpus does not clearly improve models' performance on RCA. Depending on the metric considered, it can both improve or degrade performance on lexical metrics. Semantic metrics remain largely unchanged by the incorporation of discussions. 
\end{tcolorbox}

\section{Practical Implementation of RCA Agent: A Case Study}
\label{sec:case_study}
Our evaluation of the \react\ agent in RQ1 and RQ2 does not fully capture the capability of the agent to dynamically plan and collect additional diagnostic data from team specific diagnostic services. Here, we explore these abilities of the agent by conducting a case study with \tname\, to shed light on these capabilities.

\begin{figure}[!htb]
\begin{tcolorbox}[arc=0pt,outer arc=0pt, ,nobeforeafter,colback=blue!0,boxrule=1pt,colframe=black!80,left=0mm,
  right=0mm,
  top=0mm,
  bottom=0mm,]
\textbf{Title: }[SettingDrift] Enable\censor{Sirius}AppliancePathCreation is drifted \\
\textbf{Description: } <empty>
\end{tcolorbox}
\caption{Sample Incident}\label{fig:case_study_incident_1}
\end{figure}

\subsection{Approach}
We work with \tname~ over a period of 4 weeks, primarily using unstructured discussions. We start by understanding their needs and the challenges they face with regard to RCA, followed by presenting them with the potential benefits and limitations associated with integrating an LLM based agent into their workflow. Next, we identify key diagnostic services used by the team in practice, how these services are used, and iteratively develop tools that can allow the agent to interface with these services. Lastly, we conduct demonstrations of the agent with a small set of incidents in a simple chat interface with the team to collect their feedback.  

\subsection{Knowledge Base Articles (KBAs)}
A common practice in large IT companies is to encode domain knowledge in internal knowledge base articles. In the context of incident management, these articles contain guidelines for how certain types of incidents must be diagnosed and mitigated, as well as key information about how to conduct these operations such as example database queries. At \company{}, engineers maintain a large number of KBAs for incident management. They help in standardizing operational procedures, facilitating sharing of knowledge across various teams, and onboarding new engineers. Many types of incidents, especially ones triggered by monitoring services, are tagged with relevant KBAs either automatically, or manually during triage. For these incidents, OCEs will have access to relevant KBAs the moment they start the RCA. Incidents that do not have associated KBAs typically require OCEs to spend time searching for and locating relevant KBAs before they can start RCA. We consider both of these scenarios in our case study. 

\subsection{Agent Development}

We reuse the \react\ agent from RQ1 and RQ2, but we replace the generalized tools with specialized tools that can access team specific diagnostic data. Based on discussions with \tname and preliminary experiments, we settled on the following set of tools:

\noindent \textbf{Database Query Tool:} We design and implement a tool that can be used by the agent to query databases and then analyze query results. The database framework used by the team utilizes a custom query language, which is somewhat similar to SQL. The tool design was informed by discussions with the team as well as analysis of several historical incidents experienced by the team. Based on our investigation, we settled on a design that uses two distinct components for this tool: the \textit{Query Execution Engine} and the \textit{Pandas DataFrame Query Engine}. The Query Execution Engine can be used by the agent to query the database. This requires not only the construction of the actual database query, but also knowledge of the cluster on which the database is deployed and the name of the database. This generic design gives the agent flexibility in making queries and also increases re-usability of this tool for other teams that are also using the same database platform. Once a query is successfully executed, the results returned by the database are transformed into a Pandas DataFrame and sent to the Pandas DataFrame Query Engine. The agent can then perform question-answering over the returned table using natural language queries. The Pandas DataFrame Query Engine itself consists of an LLM, which, based on the agent's queries, performs transformations on the DataFrame using the Python Interpreter and then generates a final answer.

\noindent \textbf{KBA Q/A Tool:} KBAs often contain critical information that is required to perform RCA, and are one of the most widely used resources for incident management at \company{}. For example, one of the key pieces of information required to use the Database Query Tool is the cluster address. This information is typically only available to OCEs via KBAs. To incorporate this information into the agent, we expose a question-answering tool over a set of KBAs (14 documents) provided by the team. The tool consists of a vectorstore containing chunks of KBAs, and an LLM which, given a query from the agent, uses knowledge from the retrieved KBA chunks to answer the query. If the incident in question has an associated KBA, we do not use the vectorstore and directly use it to answer questions posed by the planner.

\noindent \textbf{KBA Planning Tool:} During preliminary experiments with the team, we noticed that the eager interleaving of thoughts and actions of \react\ can be detrimental to high level planning, i.e. it can sometimes start unsuccessfully carrying out troubleshooting tasks without constructing a high level plan to guide RCA. To mitigate this phenomenon, we introduce a variant of the KBA Q/A tool which is designed to be used specifically for planning. Structurally, it is identical to the Q/A tool, but introducing it explicitly into the action space of the agent encourages it to consistently construct high level plans before taking concrete diagnostic steps. 

\noindent \textbf{Human Interaction Tool: } Our discussions with the team revealed several scenarios instances where a human-in-the-loop style workflow is necessary for RCA. For example, diagnosing certain types of incidents requires reproducing the error reported in the incident, or manually logging into a cloud device and extracting diagnostic information, which would be difficult for the agent to do. Therefore, it is desirable to have the ability for the OCE to collect such information, and provide it as an observation to the agent. Moreover, our preliminary experiments revealed that the agent struggles to make progress when key information is missing in the KBAs (such as missing cluster address for DB queries), but this information can often easily be provided by the OCE to the agent. Therefore, we add a Human Interaction Tool to allow the agent to request diagnostic information from OCEs, and also add UI enhancements to allow OCEs to interject the agent's action steps, manually verify tool executions and provide explicit feedback to the agent when desired. 
\subsection{RQ3 Results: How effective are LLM based agents at root cause analysis when given access to specialized tools used by a team for incident management?}

\subsubsection{Challenges faced by OCEs in the team for RCA}
\tname develops and maintains core services within the company's cloud platform, which hosts both internal and external customers that host cloud applications on their platform. While many of the incidents they receive are human reported, also maintain several systems that automatically detect and report error states. They maintain a large number of troubleshooting guides (KBAs) to mitigate the diversity of incident types and associated diagnostic steps. When a KBA exists for a certain type of incident, and the incident is relatively simple, new engineers with limited experience with the team's services are able to effectively perform RCA. However, identifying the right KBA for an incident can take time, and when incidents get more complex, a significant amount (> 1.5 years) of experience is required for RCA. 

\subsubsection{Real world RCA using \react}   
 We started by investigating simple incidents which have a clear KBA article available, and requires a straightforward sequence of diagnostic steps with minimal branching. We use the incident shown in Figure~\ref{fig:case_study_incident_1} as an illustrative example of incidents of this type. 
 This incident reports that there has been a \textit{setting drift} in a cluster, i.e. a setting is out of sync with the central orchestrating server. 
 This is a type of incident that is automatically reported by monitoring services, which is why the description is empty. Diagnosing this incident can lead to exactly two outcomes: 1) if there are no tenants in the affected cluster, the incident is marked as a false positive and no mitigation is required and 2) if the cluster is hosting tenants, then the OCE must identify the affected clusters (this information isn't present in the incident report) and manually instantiate a job that will rectify the setting drift to mitigate the incident. Identifying the correct outcome requires querying a database to identify affected clusters and analyzing the returned table to determine whether the incident is a false positive, or requires mitigation. Lastly, the incident report includes an associated KBA describing the necessary troubleshooting steps, example database queries as well as key pieces of information such as the database address. 
 
 Even though this incident is relatively straightforward for OCEs, it is not possible to identify whether it is a false alarm or not based only on the incident report. This underlines the importance of having access to diagnostic APIs for any automated RCA mechanism. In particular, it is worthwhile to note that even if an automated approach is able to correctly predict the outcome without carrying out the proper diagnostic steps, \textit{the OCE would still have to carry them out to verify the prediction}. When we tested \react\ agent on this incident, it is able to correctly identify case 1 consistently. This involved using the KBA Planning Tool to gather the required troubleshooting steps, adapt and execute the sample query from the KBA, and correctly assess the resulting table. While this series of action is not challenging for OCEs to execute, we stress the fact that \react\ agent has no prior knowledge of the domain, the incident or the syntax of the database query language. Yet, it is able to leverage the KBA to autonomously complete the RCA process. We observed that the agent would sometimes fail to execute the database query in its first attempt. However, since we surface appropriate error messages to the agent as observations, it was consistently able to rectify these mistakes and complete the troubleshooting process. One engineer expressed that they were "amazed by the tool's capability to automatically discern the right parameters and even rectify mistakes when the parameters are initially incorrect by querying the documents". On the other hand, the second outcome of this incident (case 2), requires an additional filtering step to remove some rows from the table returned by the database query. In our demonstrations with the team, the agent is unable to resolve this error consistently, but engineers were able to use the human-in-the-loop features of the prototype to intervene and fix the error encountered in the filtering step. 
 
 We also examined complex incidents from the team that did not have a clear set of troubleshooting steps in a single KBA, i.e. it required combining information from multiple KBAs. The diagnosis steps typically involved a series of database queries. Engineers on \tname indicated that these incidents require at least a year of experience with the team's services to effectively diagnose. Here, we observed that while the agent initially produces a plausible high level plan, it was only ever able to successfully execute one or two diagnostic steps before reaching the iteration limit (20). This is primarily due to the difficulty in producing database queries for these incidents, as information is distributed over multiple KBAs, e.g. sample queries and cluster address are not in the same KBA, requiring the agent to query the KBA Q/A tool multiple times before being able to execute a query. While the iteration limit can be extended, it will eventually fill the context. This signals the need for scaleable multi-trial framework, where experience from past trials can be used to guide future trials (e.g. \cite{Shinn_undated-kf,Zhao2023-rd,Madaan2023-tk}.

\subsection{Learnings and Practical Considerations}
In this section, we distill some key considerations for the implementation of practical LLM based agents based on our experience in the case study and feedback from OCEs.

\noindent\ul{\textbf{KBAs are critical to real world RCA}}. As seen from our findings, KBAs, are critical to performing real world RCA. They contain both specialized domain knowledge and auxillary facts about the agent's environment (e.g. database addresses, API information) that are required both for OCEs and LLM agents to effectively carry out diagnostic steps. Even experienced OCEs must either refer to KBAs in real-time or have internalized the information present in these KBAs to some degree to perform RCA. While some of this information can also be gleaned from historical incidents, incident reports typically only contain the outcome of diagnosis steps (commonly in discussion comments) rather than operational knowledge of how these steps must be performed. This is also why engineers across the company invest significant time and effort into the construction and maintenance of KBAs. 

\noindent \ul{\textbf{Tool usage in RCA is non-trivial.}} While LLMs such as \textsc{GPT-4} have shown remarkable ability to use tools prevalent in NLP such as retrieval and search, querying of diagnostic services using specialized query languages requires some trial and error. For this reason, we found that it was critical to surface error messages to the agent to provide feedback to the agent in instances of tool failures. One optimization in this regard is to replace LLMs used in tools with smaller models finetuned for tool usage. In real world settings, if we are able to scope out a set of common \textit{parameterized} services that can be specialized to different teams, finetuning the planning model to the generic usage of these tools might also significant gains.

\noindent \ul{\textbf{For complicated workflows, experiential learning and multi-trial workflows are necessary.}} Incidents that require a long and complex sequence of diagnostic steps for RCA typically have a large space of possible action trajectories. This poses significant challenges for the agent. For these incidents, single trial RCA, where we restrict the agent trajectory to 20 steps, is not sufficient. Extending the agent to a multi-trial setting necessitates the use of a reflection~\cite{Shinn2023-hb,Madaan2023-tk} or long term memory component to be able to preserve progress across trials, that allows for experiential learning. These mechanisms allow for learning based using natural language as the medium, and present opportunities for building a system where learnings from a specific team's incidents can be stored in a database, and retrieved for performing RCA on future incidents for the team. 

\noindent \textbf{\ul{Human intervention is necessary to build trust and provide some guardrails for LLMs for critical operations}}. There are many diagnostic steps which can be easily carried out by OCEs, but are not accessible as consumable services for the agent. Moreover, when agents struggle with certain parts of the diagnostic process, such as in our study, engineers can easily intervene and correct the agent's trajectory. Therefore, we recommend that agents used in practical incident management scenarios be endowed with capabilities to allow for human interaction, using a combination of explicit tools in the agent's action space, and UI features for the application exposing the agent to users. These capabilities can also be incredibly useful when combined with experiential learning; one can imagine a scenario where an engineer supervises an agent for a small set of team specific incidents, while it builds its repository of experiences, to enable quick domain adaptation for team specific knowledge, and avoid the burden of building a fine-tuning dataset for adaptation purposes.

\section{Threats to Validity}

The evaluation of the agent and other baselines for RCA are conducted on an internal dataset collected at \company , and might not apply to datasets constructed from other organizations. We use a smaller sample (n=500) of our test set to satisfy budget constraints which might not reflect performance on the larger test set. However, we minimize this threat by using random sampling, and a sample size that has been employed in prior studies. Another threat to validity comes from our manual annotations to qualitatively characterize model predictions. We mitigate this by adapting labelling criterion from prior work, and engage in multiple rounds of discussion to converge on particularly ambiguous examples. 

\section{Conclusion \& Future Work}
This work provides an empirical evaluation of an LLM-based agent, ~\react\, for root cause analysis for cloud incident management. To the best of our knowledge, this is the first empirical evaluation of LLM agents for RCA. We have shown that in an out of domain, zero-shot setting, \react\ can perform competitively with strong baselines such as retrieval augmented generation and \cot, while offering substantially lower rates of factual inaccuracies. We also showed that the use of discussion comments from incident reports does not have a significant impact on the agent's performance, revealing the limitations of performing RCA on a static dataset. Lastly, through our case study, we demonstrate the potential of LLM-agents to autonomously perform RCA in a real world setting when given access to the right tools.  
The work presented here is a first step in the development of LLM-based agents for practical RCA. One of the most promising directions for future work is the construction of a simulated RCA environment. This would overcome the limitations of a static dataset, and rapidly enhance the development of agent based approaches for RCA. As we continue to explore these avenues of future work, we anticipate that \react\ and similar agents will play a pivotal role in advancing incident management practices and automating complex decision-making processes in the software engineering domain.

\bibliography{main.bib}
\end{document}